\documentclass[%
 reprint,showpacs,
 amsmath,amssymb,
 aps,
]{revtex4-1}
\usepackage{color}
\usepackage{graphicx}
\usepackage{dcolumn}
\usepackage{bm}

\begin{document}

\preprint{APS/123-QED}

\title{The structure and dynamics of self-assembling colloidal monolayers\\ in oscillating magnetic fields}

\author{Alison E. Koser}
\author{Nathan C. Keim}%
\author{Paulo E. Arratia}
\affiliation{%
Department of Mechanical Engineering and Applied Mechanics, University of Pennsylvania, Philadelphia, USA\\}%

\date{\today}

\begin{abstract}

Many fascinating phenomena such as large-scale collective flows, enhanced fluid mixing and pattern formation have been observed in so-called active fluids, which are composed of particles that can absorb energy and dissipate it into the fluid medium. For active particles immersed in liquids, fluid-mediated viscous stresses can play an important role on the emergence of collective behavior. Here, we experimentally investigate their role in the dynamics of self-assembling magnetically-driven colloidal particles which can rapidly form organized hexagonal structures. We find that viscous stresses reduce hexagonal ordering, generate smaller clusters, and significantly decrease the rate of cluster formation, all while holding the system at constant number density. Furthermore, we show that time and length scales of cluster formation depend on the Mason number (Mn), or ratio of viscous to magnetic forces, scaling as $t\propto$ Mn and $L \propto \mbox{Mn}^{-1/2}$. Our results suggest that viscous stresses  hinder collective behavior in a self-assembling colloidal system.



\end{abstract}

\maketitle


\section{\label{sec:level1}Introduction}

Recently, there has been much interest in understanding the flow behavior and dynamics of active fluids, which arise when active or live particles are present in the fluid medium~\cite{Vicsek2012,Ramaswamy2010,Ginelli2010}. Active fluids differ from their passive counterparts in that the particles have the ability to absorb or inject energy and to generate motion and mechanical stresses in a fluid medium. Importantly, these active particles can drive the system out of equilibrium even in the absence of external forcing, as in the case of bacterial suspensions~\cite{Sokolov2007, Dombrowski2004, Mendelson1999}. Active fluids exhibit novel properties not seen in regular (passive) fluids such as large-scale flows and collective motion on length scales much greater than the particle dimensions~\cite{Sokolov2007, Dombrowski2004,Mendelson1999}, anomalous shear viscosity~\cite{Rafai2010}, giant density fluctuations~\cite{Narayan2007}, and enhanced fluid mixing~\cite{Sokolov2009}. In general, such particles interact in a plethora of ways including simple contacts in vibrated monolayers of granular particles~\cite{Narayan2007} and long-range chemical signaling among cells~\cite{BenJacob2000}. Despite the wide variety of interactions in active materials, they share a striking similarity: the emergence of rich nonlinear, collective behavior such as schooling~\cite{Vicsek2012}, clustering~\cite{Yang2008, Riedel2005}, phase segregation~\cite{Redner2013}, and pattern formation~\cite{Vicsek2012,BenJacob2000,Aranson2006}.

The role of particle interactions in the emergence of collective behavior however is not yet fully understood~\cite{Vicsek2012,Ramaswamy2010,Ginelli2010}. In particular, fluid-mediated viscous stresses are known to significantly alter the emergence of collective behavior~\cite{Goldstein2013,Jager2012_2, Yang2008,Sokolov2007}. For instance, hydrodynamic interactions can synchronize rotating paddles~\cite{Qian2009}, order colloidal crystals~\cite{Jager2012_2} and aggregate swimming sperm cells~\cite{Yang2008}. In the case of suspensions of microorganisms, large-scale collective flow manifests at sufficently high number densities but is nonexistent in dilute suspensions, leading to the conjecture that hydrodynamic interactions are the principal cause of collective flow~\cite{Sokolov2007, Riedel2005,Dombrowski2004, Mendelson1999}.

Experiments with microorganisms or other living systems, however, can be sensitive to factors such as temperature or chemical conditions, and the relative contribution of viscous stresses specifically hydrodynamic interactions is typically a function of number density and not an independent variable. Furthermore, interactions and forcing among microogranisms can be complex and difficult to quantify. Thus, there is a need for nonliving experimental models which can withstand large variations of independent parameters for testing hypotheses involving active fluids. Examples include vibrated granular materials~\cite{Narayan2007}, synthetic photoactivated colloids~\cite{Palacci2013}, and chemically reactive colloidal particles~\cite{Takagi2013}. Another model system is a suspension of paramagnetic particles which can be activated via external magnetic fields and dynamically self-assemble~\cite{Tierno2007}. This is an interesting overdamped system for exploring whether viscous stresses and hydrodynamic interactions amplify or hinder the cluster formation process. 

Paramagnetic and magnetic particles are known to exhibit collective behavior. In steady or slowly rotating magnetic fields such particles form chains, which align with the magnetic field lines~\cite{Petousis2007, Biswal2004, Vuppu2003,Furst1999,Helgesen1988}. If the field rotates too quickly, viscous drag breaks apart 3he chains~\cite{Petousis2007}, and the particles can instead form highly organized crystals~\cite{Jager2012, Regtmeier2012, Wittbracht2011}. However, previous experiments have not yet explored the role of hydrodynamic interactions in cluster formation.

In this manuscript, we investigate the role of fluid-mediated viscous stresses on the structure and dynamics of self-assembling paramagnetic particles. We find that such stresses hinder the cluster formation process. In this active material, viscous stresses, which are varied independently from density, (1) reduce the hexagonal order of clusters, (2) decrease their size, and (3) significantly slow down their rate of formation.

\begin{figure}
\includegraphics[width=.9 \linewidth]{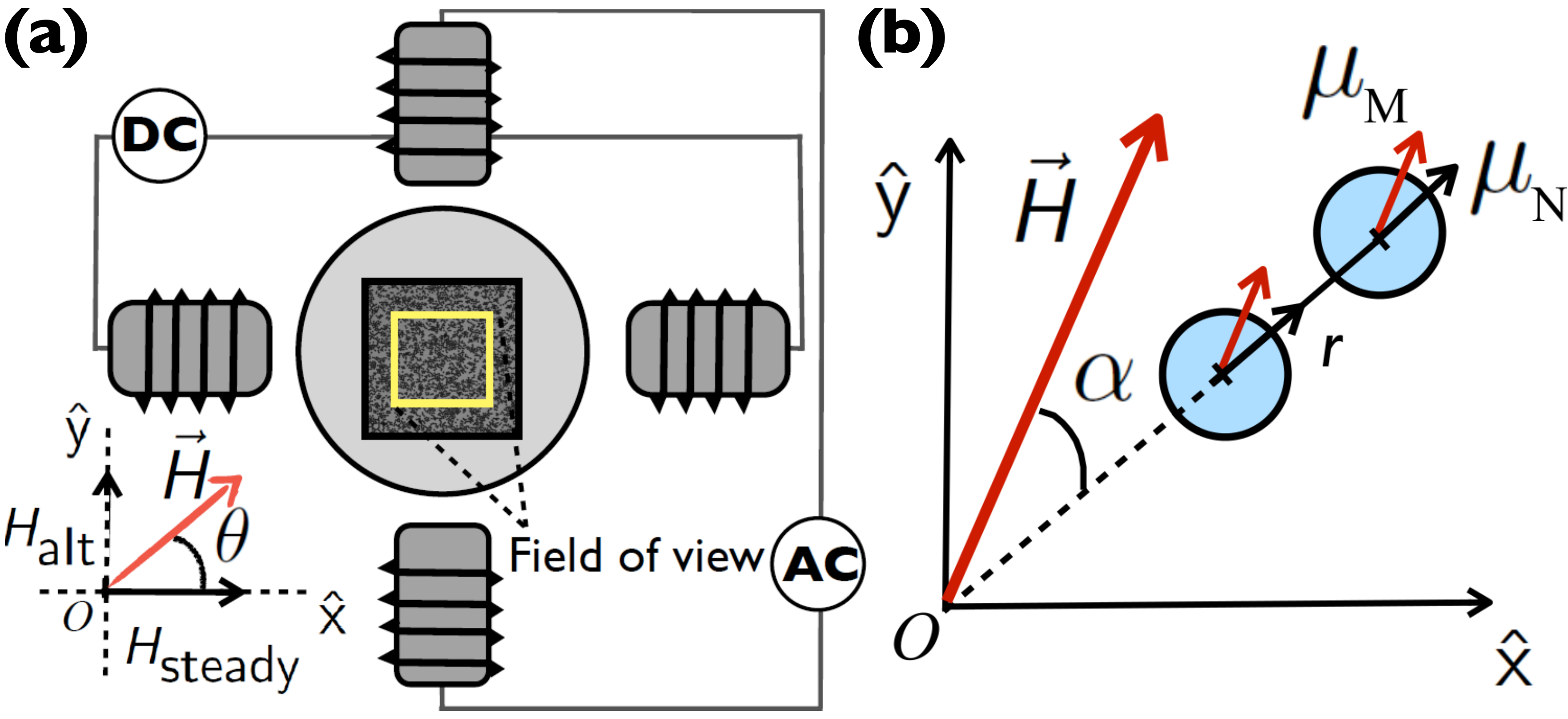}
\caption{\label{Fig1}  (a) Experimental setup. Four computer-controlled electromagnets create an oscillating magnetic field in the plane of the particles. (b) Magnetic forces between two paramagnetic particles in an external magnetic field. Each particle has a moment vector $\mu_M$ in the direction of the applied magnetic field and one moment vector $\mu_N$ along the axis connecting the centers of each sphere. }
\end{figure}

\section{\label{sec:level1}Experimental Methods}

Experiments are performed with a monolayer of spherical paramagnetic particles suspended in water and activated by an oscillating external magnetic field. The particles (Spherotech) are 20.5 $\mu$m in diameter, stabilized with a carboxyl-group coating, and have an effective magnetic susceptibility $\chi = 0.06$ \cite{Derks2010}. The monolayer is generated by confining an aqueous suspension of the particles in a small cell, comprising PDMS side walls and sealed above and below with glass slides. The cell is 6 mm by 6 mm in area and 1 mm in height. Since the particles are relatively large, gravity quickly settles them into a monolayer at the bottom of the cell. The particles stay a few nanometers above the glass surface due to the electrostatic repulsion between the carboxyl-group coating and the glass surface \cite{Tierno2007}. Here, the area fraction $\phi$ is $0.42 \pm 0.02$. This area fraction was chosen so that the monolayer is dense enough to form clusters but still far below the jamming transition at 0.84.

To observe dynamical assembly, the particles are placed in an oscillating magnetic field $\vec{H}(t)$  generated by four computer-controlled electromagnets, as shown in \textbf{Fig.~\ref{Fig1}a}. Two of the electromagnets supply a steady field of magnitude $H_{\text{steady}}$. The other two electromagnets, orthogonal to the first pair, supply an alternating field, a sinusoidal wave of amplitude $H_{\text{alt}}$ and held at a constant frequency $f$ equal to 0.2 Hz. The combined magnetic field can be written as:

\begin{equation}
\vec{H}(t) =H_{\text{steady}} \hat{\mbox{x}}+ H_{\text{alt}}\mbox{sin}(\omega t) \hat{\text{y}},
\label{H}
\end{equation}

\noindent where $\omega=2\pi f$ is the angular frequency. 

\begin{figure}
\includegraphics[width=.8 \linewidth]{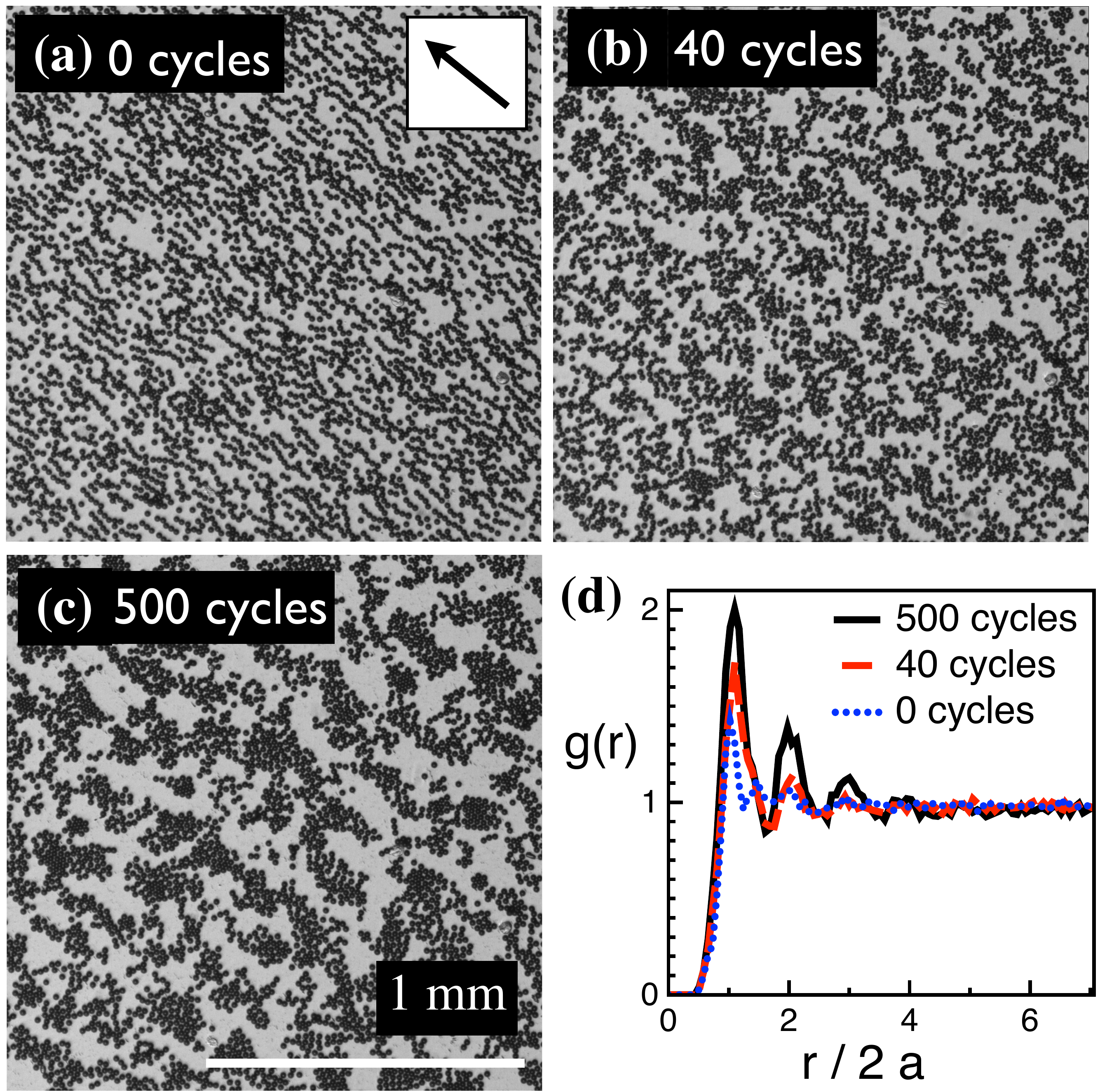}
\caption{\label{Fig2} Formation of clusters with paramagnetic particles. Snapshots of experiment conducted at $\mbox{Mn} = 6.8\times10^{-4}$ and $\phi = 0.41$ during (a) its initial configuration, (b) after 40 cycles, and (c) after 500 cycles. In (a), the arrow denotes the direction of the aligning field $H_{\mbox{steady}}$ for all images shown in this paper. (d) The corresponding pair correlation functions $g(r)$ for each snapshot. The first three peaks of $g(r)$ increase as the number of cycles increases.}
\end{figure}

\begin{figure*}
\includegraphics[width=.9 \linewidth]{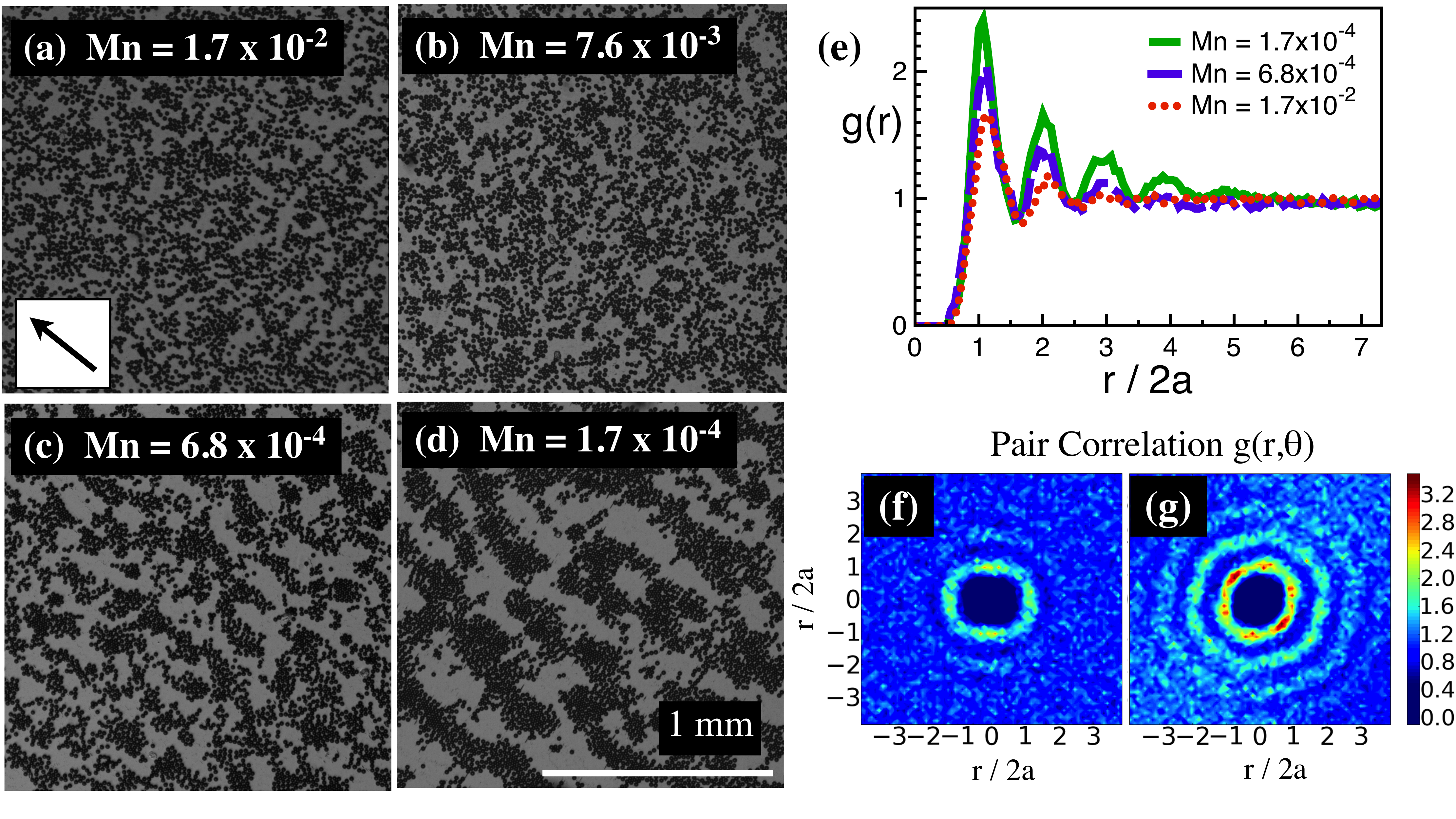}
\caption{\label{Fig3} (a-d) Sample snapshots of cluster formation as a function of Mason number (Mn).  The arrow denotes the direction of the aligning field $H_{\mbox{steady}}$ for all images shown in this paper. (e) Pair correlation function $g(r)$ after 500 cycles as a function of Mn. The first four peaks increase as Mn is lowered from $1.7\times 10^{-2}$ to $1.7\times 10^{-4}$. (f-g) The magnitude of the pair correlation vector $g(r,\theta)$ for (f) $\mbox{Mn} =1.7\times 10^{-2}$ and (g) $\mbox{Mn} = 1.7\times 10^{-4}$. Note that viscous stresses are more relevant as Mn increases. The six peaks at $r/2a=1.0$ indicate hexagonal ordering when magnetic forcing is strongest.}
\end{figure*}

We note that this experimental design is different from previous experiments involving monolayers of dipolar particles in rotating electric or magnetic fields~\cite{Regtmeier2012,Wittbracht2011,Tierno2007,Snoswell2006,Vuppu2003} in the sense that one component of the magnetic field remains steady. This introduces a new degree of freedom, the angle $\theta$ through which the magnetic field sweeps, as sketched in \textbf{Fig.~\ref{Fig1}}. The angle $\theta$ is defined by the ratio of the two field strengths, $\theta = \mbox{tan}^{-1}(H_{\text{alt}}/H_{\text{steady}})$. In the limit of small $H_{\text{alt}}/H_{\text{steady}}$, the field approaches the steady case in which the particles form one-dimensional chains~\cite{Vuppu2003, Furst1999, Helgesen1988} and no two-dimensional aggregates would form. Previous work has shown that the break up of chains and formation of clusters occurs when a critical frequency~\cite{Vuppu2003, Petousis2007} or a critical angle between the field and the plane of the particles~\cite{Tierno2007} is reached. Likewise, we expect a critical angle for this new degree of freedom ($\theta$) which must also be exceeded for chains to break and particles to cluster. Here, we wish to investigate the role of viscous stresses in cluster formation. Therefore, we choose an angle in which we observe clustering, $\theta = 75 \,^{\circ}$ and hold it fixed, leaving the investigation of $\theta$ on the clustering mechanism for later experiments.

Due to the paramagnetic nature of the particles, when placed in the field $\vec{H}$, interparticle magnetic forces arise via induced dipole moments $\mu$ in the particles. The dipole moment $\mu= 4/3 \pi a^3 \mu_0 \chi H$ is proportional to and aligns with the magnetic field lines. The dipole moment is dependent on the particle radius \emph{a}, the magnetic permeability of a vacuum $\mu_0$, and the particle susceptibility $\chi$. Two particles in the field experience a dipole-dipole force due to the magnetic field generated by each of their moments $\mu$. This force is given by~\cite{Biswal2004}:

\begin{equation}
\vec{F}_{\mbox{mag}}=\frac{3\mu^2}{4\pi \mu_0 r^4}(3\mbox{cos}^2 \alpha -1)\hat{r} + \frac{3 \mu^2}{4 \pi \mu_0 r^4} \mbox{sin}(2\alpha) \hat{\theta},
\label{MagF}
\end{equation}

\noindent where $r$ is the center-to-center distance between the particles and $\alpha$ is the angle between $\vec{r}$ and $\vec{H}$, shown in \textbf{Fig.~\ref{Fig1}b}.

The relative contribution of the attractive magnetic forcing $f_m \sim \mu_0 a^2 \chi^2 H^2$ to the viscous drag $f_d \sim \eta a v$ experienced by the particle is usually described by the Mason number Mn, defined as:

\begin{equation}
\mbox{Mn} = \frac{32 \eta \omega}{\mu_0 \chi^2 H^2},
\label{Mn}
\end{equation}

\noindent{}where $\omega$ is the frequency of the external magnetic field, $\eta$ is the viscosity of the fluid, and the velocity of the particles is assumed to be $a\omega$. The prefactor 32 is derived from the torque balance on a chain of particles~\cite{Biswal2004} and is used here because of the initial chain formation as seen in \textbf{Fig.~\ref{Fig2}a}. For particles to move and cluster, the magnetic forces on a particle must overcome viscous drag, i.e. $\mbox{Mn} <1$. The Mason number is varied via the field strength $H$ from $6.8 \times 10^{-2}$ to $1.7 \times 10^{-4}$.

\begin{figure*}
\includegraphics[width=.75 \linewidth]{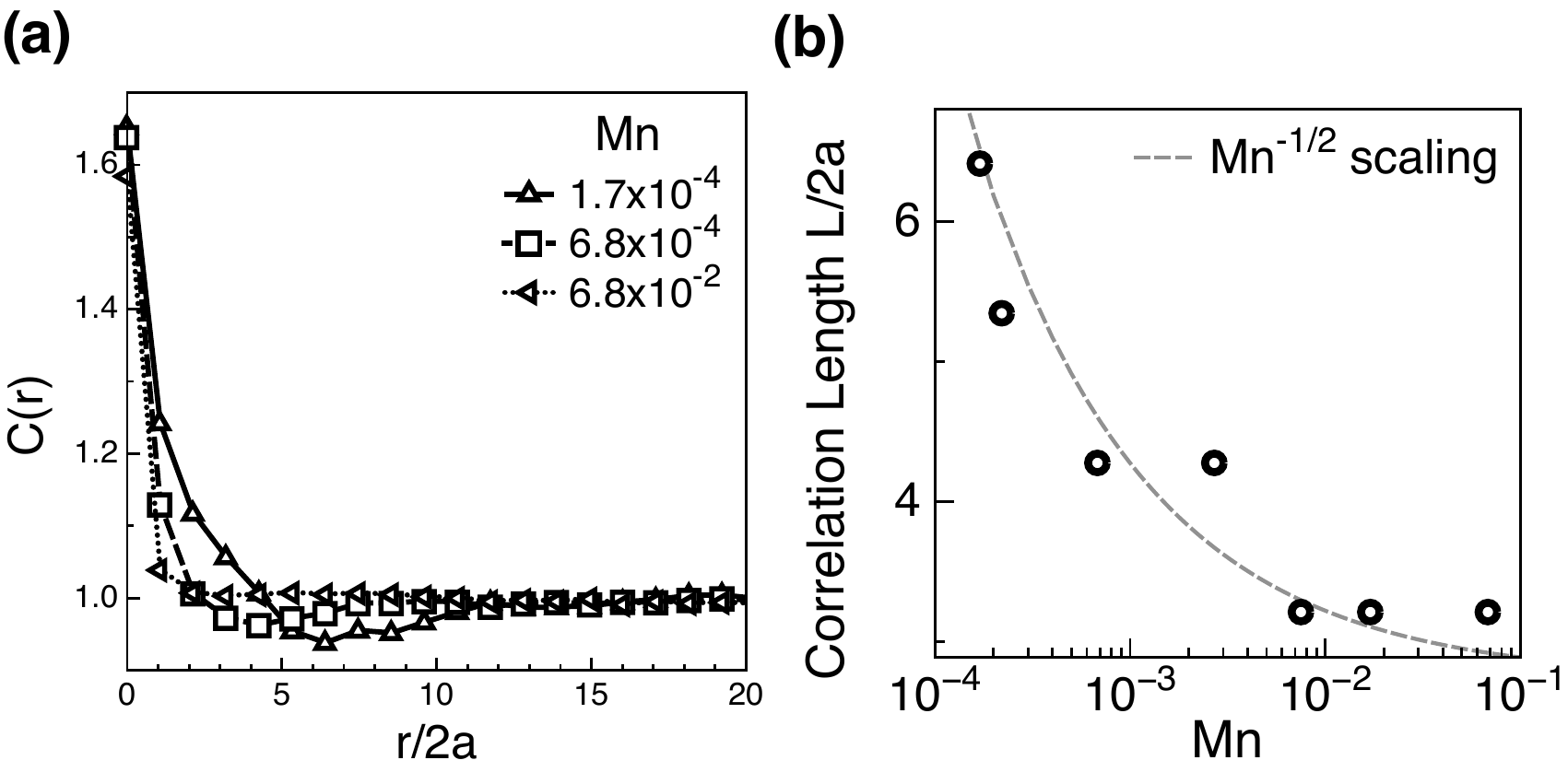}
\caption{\label{Fig4} (a) Coarse-grained density autocorrelation function at $\mbox{Mn} = 1.7 \times 10^{-4}$, $6.8 \times 10^{-4}$ and $6.8 \times 10^{-2}$. The minimum in the density autocorrelation function, or the correlation length, shifts to the left as Mn is raised. (b) Correlation length versus Mn. The correlation length decreases from 6.4 to 3.2 particle diameters as Mn increases, indicating smaller clusters when viscous stresses are significant. The dotted line represents $L\propto \mbox{Mn}^{-1/2}$ scaling.}
\end{figure*}

\section{\label{sec:level1}Results and Discussion}

Initially, the particles are positioned within a steady magnetic field. As has been previously shown~\cite{Vuppu2003,Furst1999,Helgesen1988}, the particles in a steady field form one-dimensional chains which align with the field. When the magnetic field begins to oscillate, the particle configurations change over time~\cite{SuppMat}.  The particles form clusters of a particular length scale. This is illustrated in \textbf{Fig.~\ref{Fig2}a-c} by snapshots after 0, 40, and 500 cycles of an experiment conducted at a single Mason number, $\mbox{Mn} = 6.8\times 10^{-2}$. 

In order to gain further insights into the structure and dynamics of the cluster formation process, the particles are tracked individually for 500 cycles. We quantify these observations as a function of time with the pair correlation function $g(r)$. \textbf{Figure~\ref{Fig2}d} shows $g(r)$ for each snapshot in \textbf{Fig~\ref{Fig2}a-c}. We find an increase in the first three peaks of $g(r)$ as the number of cycles increases. This is a result of the particles gaining more neighbors and clustering over longer length scales in comparison to their initial configuration.

Next, we vary the Mason number to study the role of viscous stresses on the structure and dynamics of cluster formation. The Mason number is adjusted via the magnetic field $H$, defined by the root-mean-square strength of the field $| \vec{H}|$ (Eqn.~\ref{H}) which ranges from 5 to 120 kA/m. As Mn is decreased, we find the particles form increasingly larger and more organized clusters~\cite{SuppMat}. This effect is illustrated in \textbf{Fig.~\ref{Fig3}} which shows images from experiments conducted at $\mbox{Mn} = 1.7 \times 10^{-2}$, $7.6 \times 10^{-3}$, $6.8 \times 10^{-4}$, and $1.7 \times 10^{-4}$ after 500 cycles. At relatively high Mn, as shown in \textbf{Fig.~\ref{Fig3}a}, viscous forces dominate and prevent particles from significant clustering and exhibiting large-scale ordering. However, as Mn is decreased (\textbf{Fig.~\ref{Fig3}d}) magnetic forces are relatively strong, and the particles form blue more organized clusters. Furthermore, \textbf{Fig.~\ref{Fig3}a-d} show that the length scale of the clusters increases as Mn is lowered.

\subsection{\label{sec:level2}Viscous stresses reduce hexagonal structure}

\textbf{Figure~\ref{Fig3}a} shows that when viscous stresses are most significant, i.e. $\mbox{Mn} = 1.7 \times 10^{-2}$, very little clustering is observed after 500 cycles. But as Mn decreases and viscous stresses become less significant, the particles tend to form blue larger crystals with nearest-neighbor hexagonal order. We quantify this observation by computing the corresponding pair correlation functions $g(r)$, as shown in \textbf{Fig.~\ref{Fig3}e}, for images in \textbf{Fig.~\ref{Fig3}a-d}. For highest Mn, there is only two distinct peaks. But as Mn is decreased, four peaks emerge, all increased in value, indicating a higher probability of finding particles at increased distances from a reference particle. This suggests that significant viscous stresses and their corresponding hydrodynamic interactions reduce long-range order.

The 2D pair correlation $g(r,\theta)$ shown in \textbf{Fig.~\ref{Fig3}f-g} also reveals details about the material structure. At low Mn, i.e.  $\mbox{Mn} = 1.7 \times 10^{-4}$ (\textbf{Fig.~\ref{Fig3}g}), $g(r,\theta)$ resembles a nearest-neighbor hexagonal structure with six distinct peaks observed around $r/2a=1$. There are deviations from a hexagonal lattice structure however. For instance, the six peaks are not of constant magnitude nor are they strictly 60$^{\circ}$ apart, in a way that may depend on our choice of $\theta$. The two highest peaks reflect the static field alignment. In contrast, for the case of higher Mn ($\mbox{Mn} = 1.7 \times 10^{-2}$), we do not observe the hexagonal pattern. The value at $r/2a=1$ is nearly uniform, and the extrema values are lower than those presented for $\mbox{Mn} = 1.7 \times 10^{-4}$, indicating fewer average number of neighbors. These measurements suggest that viscous stresses reduce the hexagonal order of clusters.

\subsection{\label{sec:level2}Viscous stresses result in smaller clusters}

Our results so far show that the particles tend to form much larger clusters at low Mn, when viscous stresses are relatively weak. To quantify a correlation length scale that reflects the size of the clusters, we consider a density-density correlation function $C(r)$. Since we are only interested in which regions contain more or fewer particles, we consider a density correlation function that is coarse-grained at a length scale above a particle diameter. To compute $C(r)$, we first measure a coarse-grained density field $\rho$ with unit cells of width $3 a$, where $a$ is the radius of a particle. Then, the autocorrelation function of $\rho$ is calculated as:

\begin{equation}
 C(r) = \frac{\langle \rho(\vec{r}_0) \rho(\vec{r}_1)\rangle}{{\langle \rho(\vec{r}_0) \rangle}^2},
\end{equation}

\noindent where $\| \vec{r}_1-\vec{r}_0 \|=r$ and the brackets denote an average over $\vec{r}_0$ and $\vec{r}_1$. Similar density-density correlation functions have been used to characterize structure~\cite{Dinsmore2001} and determine fractal dimensions of colloidal aggregates~\cite{Gonzalez1999}.

\begin{figure*}
\includegraphics[width=.8 \linewidth]{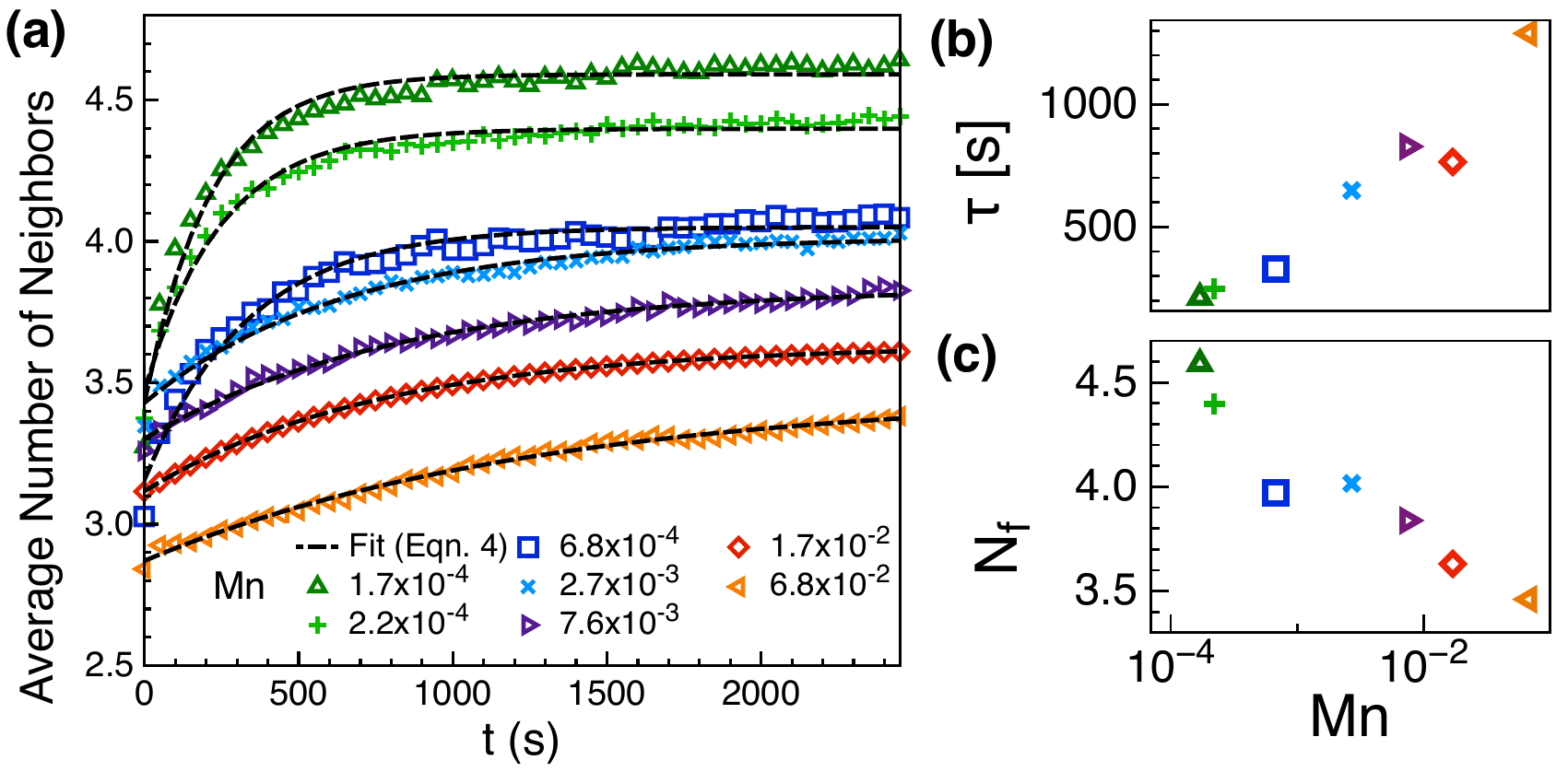}
\caption{\label{Fig5} (a) The average number of neighbors as a function of time and Mn. (b) Timescale as a function of Mn. The clustering time $\tau$ increases with Mn indicating that viscous stresses slow down the dynamics. (c) The final average number of neighbors versus Mn, which also decreases with Mn.}
\end{figure*}

\textbf{Figure~\ref{Fig4}a} show the values of $C(r)$ as a function of $r/2a$ for $\mbox{Mn} =1.7 \times 10^{-4}$, $6.8 \times 10^{-4}$, and $6.8 \times 10^{-2}$. The data suggests that cluster formation leads to correlations within a cluster and anticorrelations outside, corresponding to a peak and minimum in $C(r)$. At small $r$ (or within a cluster), the local densities are correlated and $C(r)$ is relatively high. As $r$ increases and extends past the length scale of a cluster, the curves decay and reach a minimum value. This minimum corresponds to decreased correlations in regions inbetween clusters where particle densities tend to be lower. We take the distance at which this minimum occurs to be the correlation length $L$ of the clusters. At large $r$, $C(r)$ approaches a constant value, indicating a uniform density correlation. 
 
\textbf{Figure~\ref{Fig4}b} displays the correlation lengths $L$ as a function of Mn. We find that $L$ decreases from 6.4 to 3.2 particle diameters. These length scales are consistent with the cluster sizes shown in \textbf{Fig.~\ref{Fig3}a-d} and the trend indicates that hydrodynamic interactions tend to decrease the size of a cluster.

A series of theoretical and experimental studies have shown that the length scales of colloidal chains in magnetic fields scale as $\mbox{Mn}^{-1/2}$ for 1D chains~\cite{Wittbracht2012,Petousis2007}; however, this has not been measured for two-dimenional clusters. The one-dimensional studies were conducted via chains of paramagnetic particles in slowly rotating fields. As a magnetically-formed chain rotates through a fluid, viscous drag, which depends on the length of the chain, acts to break it apart. Petousis \emph{et al.} showed there is a maximum length under which a chain is stable for a given Mn which scales as $\mbox{Mn}^{-1/2}$~\cite{Petousis2007}. The unstable regime in which chains break apart and form two-dimensional clusters has only recently been explored~\cite{Regtmeier2012,Wittbracht2012,Wittbracht2011}. Here, we present the correlation length of two-dimensional clusters $L$ as a function of Mn. \textbf{Figure ~\ref{Fig4}b} shows that the correlation length is consistent with $L\propto \mbox{Mn}^{-1/2}$, as represented by the dotted line and which matches previous studies of stable one-dimensional chain lengths.

\subsection{\label{sec:level2}Viscous stresses slow clustering dynamics}

To quantify the clustering dynamics, we measure the average number of neighbors $N$ over time $t$ for a range of Mason numbers. Neighbors are defined as particles whose center-to-center distance is less than the distance at which the first minimum in $g(r)$ occurs. We use the average number of neighbors $N$ as a measure of dynamics instead of the correlation length $L$ since $N$ can be measured at initial times before a correlation length is apparent. \textbf{Figure~\ref{Fig5}a} shows that the average number of neighbors increases over time for all Mn; however, the rate at which it increases varies. In order to determine a characteristic clustering time $\tau$, we fit the data in \textbf{Fig.~\ref{Fig5}a} to the form

\begin{equation}
N = N_0+(N_f-N_0)\bigg(1-e^{-t/\tau} \bigg),
\label{Neighbors}
\end{equation}

\noindent where $N_0$ and $N_f$ are the initial average number of neighbors and the final average number of neighbors respectively. \textbf{Figure~\ref{Fig5}b} and \textbf{c} show that the time scale $\tau$ increases from 212 to 1290 s and the final number of neighbors $N_f$ decreases from 4.6 to  3.5. Thus, increases in viscous stresses (or increases in Mn) tend to increase the clustering time and lower the hexagonal order farther from its limit of $N_f = 6$ at the highest possible packing fraction.

Calderon and Melle~\cite{Calderon2002} showed that the dynamics of paramagnetic particles within rotating magnetic fields should depend on a temporal scale $t_s$,

\begin{equation}
t_s= \mbox{Mn} /\omega
\label{t}
\end{equation}

\noindent{}where $\omega$ is the angular frequency of the applied field. This implies that for lower Mn (relatively stronger magnetic forcing) the dimensionless time scale should decrease and faster cluster formation is expected.

\begin{figure}
\includegraphics[width=.85 \linewidth]{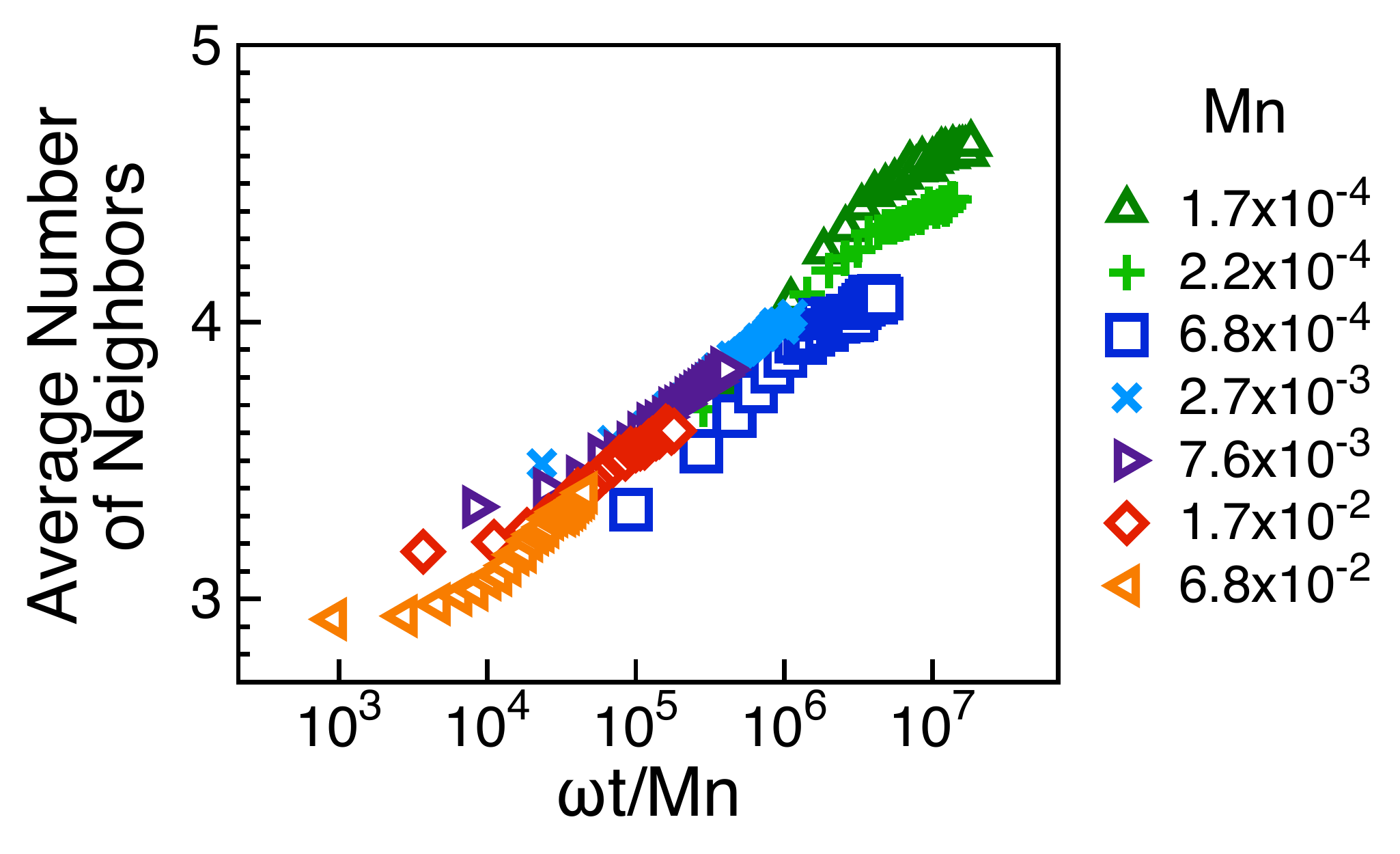}
\caption{\label{Fig6} Average number of neighbors versus $\omega t /\mbox{Mn}$. The data collapse over many decades in Mn, reinforcing the time scaling $t \propto \mbox{Mn}$ also used in~\cite{Calderon2002}.}
\end{figure}

To illustrate this time scaling, \textbf{Fig.~\ref{Fig6}} shows the average number of neighbors versus a rescaled time, $\omega t/\mbox{Mn}$. The data aligns over four orders of magnitude. Note, we do not expect perfect alignment. When the magnetic field is stronger, more particles contribute to one-dimensional aggregates, raising the initial number of neighbors. Furthermore, each curve deviates at the end as a result of the saturation in final average number of neighbors $N_f$. Despite these differences, there is still an overall agreement in the trend of the master curve, consistent with $t_s \propto \mbox{Mn}/\omega$.

Previously, the clustering rate for magnetic particles in a rotating magnetic field was measured for three different frequencies by Wittbracht \emph{et al.}~\cite{Wittbracht2011}. The investigators found that the clustering rate increased linearly with frequency which is also consistent with the scaling of Calderon and Melle (Eqn.~\ref{t})~\cite{Calderon2002}. Here, we have showed the validity of Eqn.~\ref{t} over a large range in Mason number, which should reflect many other magnetorheological fluids with different geometries or particles. We anticipate that the role of fluid-mediated viscous stresses on the structure and dynamics presented here can be extended to other active fluids.

\section{\label{sec:level1} Conclusion}

In this paper, we investigated the role of viscous stresses on the structure and dynamics of cluster formation in monolayers of paramagnetic particles subjected to an oscillating magnetic field. When viscous forcing is relatively low, the particles form large clusters with nearest-neighbor hexagonal ordering. But as the role of viscous stresses and hydrodynamic interactions increase, the structure becomes less ordered and the clusters become smaller, their length scaling as $L \propto \mbox{Mn}^{-1/2}$. At the same time, viscous stresses slow cluster formation. The clustering time increases significantly from 212 to 1290 s as Mn increases. This is consistent with the presence of a time scale $t_s = \mbox{Mn}/\omega$, as shown by the collapse of the average number of neighbors versus $\omega t/\mbox{Mn}$, ranging over several orders of magnitude. These results are in contrast to the conventional view that viscous stress, in particular hydrodynamic interactions, accelerate and/or cause collective behavior. Furthermore, the role of viscous stresses in (1) reducing cluster size, (2) creating less order and (3) slowing dynamics may have significant impact on the collective behavior in other active fluids, such as the clustering of \emph{E. coli} bacteria~\cite{Tailleur2008}, aggregation of sperm cells~\cite{Yang2008}, or synchronization of rotating flagella~\cite{Reichert2005}.

\begin{acknowledgments}

We thank Gabriel Redner for helpful discussions and Mike Garcia for early experimental work. This work is supported by the Army Research Office through award W911NF-11-1-0488.

\end{acknowledgments}

\bibliography{active_colloids_ref2}

\providecommand{\noopsort}[1]{}\providecommand{\singleletter}[1]{#1}%
\begin{thebibliography}{37}%
\makeatletter
\providecommand \@ifxundefined [1]{%
 \@ifx{#1\undefined}
}%
\providecommand \@ifnum [1]{%
 \ifnum #1\expandafter \@firstoftwo
 \else \expandafter \@secondoftwo
 \fi
}%
\providecommand \@ifx [1]{%
 \ifx #1\expandafter \@firstoftwo
 \else \expandafter \@secondoftwo
 \fi
}%
\providecommand \natexlab [1]{#1}%
\providecommand \enquote  [1]{``#1''}%
\providecommand \bibnamefont  [1]{#1}%
\providecommand \bibfnamefont [1]{#1}%
\providecommand \citenamefont [1]{#1}%
\providecommand \href@noop [0]{\@secondoftwo}%
\providecommand \href [0]{\begingroup \@sanitize@url \@href}%
\providecommand \@href[1]{\@@startlink{#1}\@@href}%
\providecommand \@@href[1]{\endgroup#1\@@endlink}%
\providecommand \@sanitize@url [0]{\catcode `\\12\catcode `\$12\catcode
  `\&12\catcode `\#12\catcode `\^12\catcode `\_12\catcode `\%12\relax}%
\providecommand \@@startlink[1]{}%
\providecommand \@@endlink[0]{}%
\providecommand \url  [0]{\begingroup\@sanitize@url \@url }%
\providecommand \@url [1]{\endgroup\@href {#1}{\urlprefix }}%
\providecommand \urlprefix  [0]{URL }%
\providecommand \Eprint [0]{\href }%
\providecommand \doibase [0]{http://dx.doi.org/}%
\providecommand \selectlanguage [0]{\@gobble}%
\providecommand \bibinfo  [0]{\@secondoftwo}%
\providecommand \bibfield  [0]{\@secondoftwo}%
\providecommand \translation [1]{[#1]}%
\providecommand \BibitemOpen [0]{}%
\providecommand \bibitemStop [0]{}%
\providecommand \bibitemNoStop [0]{.\EOS\space}%
\providecommand \EOS [0]{\spacefactor3000\relax}%
\providecommand \BibitemShut  [1]{\csname bibitem#1\endcsname}%
\let\auto@bib@innerbib\@empty
\bibitem [{\citenamefont {Vicsek}\ and\ \citenamefont
  {Zafeiris}(2012)}]{Vicsek2012}%
  \BibitemOpen
  \bibfield  {author} {\bibinfo {author} {\bibfnamefont {T.}~\bibnamefont
  {Vicsek}}\ and\ \bibinfo {author} {\bibfnamefont {A.}~\bibnamefont
  {Zafeiris}},\ }\href@noop {} {\bibfield  {journal} {\bibinfo  {journal}
  {Phys. Reports}\ }\textbf {\bibinfo {volume} {517}},\ \bibinfo {pages} {71}
  (\bibinfo {year} {2012})}\BibitemShut {NoStop}%
\bibitem [{\citenamefont {Ramaswamy}(2010)}]{Ramaswamy2010}%
  \BibitemOpen
  \bibfield  {author} {\bibinfo {author} {\bibfnamefont {S.}~\bibnamefont
  {Ramaswamy}},\ }\href@noop {} {\bibfield  {journal} {\bibinfo  {journal}
  {Annu. Rev. Condes. Matter Phys.}\ }\textbf {\bibinfo {volume} {1}},\
  \bibinfo {pages} {323} (\bibinfo {year} {2010})}\BibitemShut {NoStop}%
\bibitem [{\citenamefont {Ginelli}\ \emph {et~al.}(2010)\citenamefont
  {Ginelli}, \citenamefont {Pernuani}, \citenamefont {B{\"a}r},\ and\
  \citenamefont {Chat{\'e}}}]{Ginelli2010}%
  \BibitemOpen
  \bibfield  {author} {\bibinfo {author} {\bibfnamefont {F.}~\bibnamefont
  {Ginelli}}, \bibinfo {author} {\bibfnamefont {F.}~\bibnamefont {Pernuani}},
  \bibinfo {author} {\bibfnamefont {M.}~\bibnamefont {B{\"a}r}}, \ and\
  \bibinfo {author} {\bibfnamefont {H.}~\bibnamefont {Chat{\'e}}},\ }\href@noop
  {} {\bibfield  {journal} {\bibinfo  {journal} {Phys. Rev. Lett.}\ }\textbf
  {\bibinfo {volume} {104}},\ \bibinfo {pages} {184502} (\bibinfo {year}
  {2010})}\BibitemShut {NoStop}%
\bibitem [{\citenamefont {Sokolov}\ \emph {et~al.}(2007)\citenamefont
  {Sokolov}, \citenamefont {Aranson}, \citenamefont {Kessler},\ and\
  \citenamefont {Goldstein}}]{Sokolov2007}%
  \BibitemOpen
  \bibfield  {author} {\bibinfo {author} {\bibfnamefont {A.}~\bibnamefont
  {Sokolov}}, \bibinfo {author} {\bibfnamefont {I.~S.}\ \bibnamefont
  {Aranson}}, \bibinfo {author} {\bibfnamefont {J.~O.}\ \bibnamefont
  {Kessler}}, \ and\ \bibinfo {author} {\bibfnamefont {R.~E.}\ \bibnamefont
  {Goldstein}},\ }\href@noop {} {\bibfield  {journal} {\bibinfo  {journal}
  {Phys. Rev. Lett.}\ }\textbf {\bibinfo {volume} {98}},\ \bibinfo {pages}
  {158102} (\bibinfo {year} {2007})}\BibitemShut {NoStop}%
\bibitem [{\citenamefont {Dombrowski}\ \emph {et~al.}(2004)\citenamefont
  {Dombrowski}, \citenamefont {Cisneros}, \citenamefont {Chatkaew},
  \citenamefont {Goldstein},\ and\ \citenamefont {Kessler}}]{Dombrowski2004}%
  \BibitemOpen
  \bibfield  {author} {\bibinfo {author} {\bibfnamefont {C.}~\bibnamefont
  {Dombrowski}}, \bibinfo {author} {\bibfnamefont {L.}~\bibnamefont
  {Cisneros}}, \bibinfo {author} {\bibfnamefont {S.}~\bibnamefont {Chatkaew}},
  \bibinfo {author} {\bibfnamefont {R.~E.}\ \bibnamefont {Goldstein}}, \ and\
  \bibinfo {author} {\bibfnamefont {J.~O.}\ \bibnamefont {Kessler}},\
  }\href@noop {} {\bibfield  {journal} {\bibinfo  {journal} {Phys. Rev. Lett.}\
  }\textbf {\bibinfo {volume} {93}},\ \bibinfo {pages} {098103} (\bibinfo
  {year} {2004})}\BibitemShut {NoStop}%
\bibitem [{\citenamefont {Mendelson}\ \emph {et~al.}(1999)\citenamefont
  {Mendelson}, \citenamefont {Bourque}, \citenamefont {Wilkening},
  \citenamefont {Anderson},\ and\ \citenamefont {Watkins}}]{Mendelson1999}%
  \BibitemOpen
  \bibfield  {author} {\bibinfo {author} {\bibfnamefont {N.~H.}\ \bibnamefont
  {Mendelson}}, \bibinfo {author} {\bibfnamefont {A.}~\bibnamefont {Bourque}},
  \bibinfo {author} {\bibfnamefont {K.}~\bibnamefont {Wilkening}}, \bibinfo
  {author} {\bibfnamefont {K.~R.}\ \bibnamefont {Anderson}}, \ and\ \bibinfo
  {author} {\bibfnamefont {J.~C.}\ \bibnamefont {Watkins}},\ }\href@noop {}
  {\bibfield  {journal} {\bibinfo  {journal} {J. Bacteriol.}\ }\textbf
  {\bibinfo {volume} {181}},\ \bibinfo {pages} {600} (\bibinfo {year}
  {1999})}\BibitemShut {NoStop}%
\bibitem [{\citenamefont {Rafa{\"\i}}\ \emph {et~al.}(2010)\citenamefont
  {Rafa{\"\i}}, \citenamefont {Peyla},\ and\ \citenamefont
  {Jibuti}}]{Rafai2010}%
  \BibitemOpen
  \bibfield  {author} {\bibinfo {author} {\bibfnamefont {S.}~\bibnamefont
  {Rafa{\"\i}}}, \bibinfo {author} {\bibfnamefont {P.}~\bibnamefont {Peyla}}, \
  and\ \bibinfo {author} {\bibfnamefont {L.}~\bibnamefont {Jibuti}},\
  }\href@noop {} {\bibfield  {journal} {\bibinfo  {journal} {Phys. Rev. Lett.}\
  }\textbf {\bibinfo {volume} {104}},\ \bibinfo {pages} {098102} (\bibinfo
  {year} {2010})}\BibitemShut {NoStop}%
\bibitem [{\citenamefont {Narayan}\ \emph {et~al.}(2007)\citenamefont
  {Narayan}, \citenamefont {Ramasway},\ and\ \citenamefont
  {Menon}}]{Narayan2007}%
  \BibitemOpen
  \bibfield  {author} {\bibinfo {author} {\bibfnamefont {V.}~\bibnamefont
  {Narayan}}, \bibinfo {author} {\bibfnamefont {S.}~\bibnamefont {Ramasway}}, \
  and\ \bibinfo {author} {\bibfnamefont {N.}~\bibnamefont {Menon}},\
  }\href@noop {} {\bibfield  {journal} {\bibinfo  {journal} {Science}\ }\textbf
  {\bibinfo {volume} {317}},\ \bibinfo {pages} {105} (\bibinfo {year}
  {2007})}\BibitemShut {NoStop}%
\bibitem [{\citenamefont {Sokolov}\ \emph {et~al.}(2009)\citenamefont
  {Sokolov}, \citenamefont {Goldstein}, \citenamefont {Feldchtein},\ and\
  \citenamefont {Aranson}}]{Sokolov2009}%
  \BibitemOpen
  \bibfield  {author} {\bibinfo {author} {\bibfnamefont {A.}~\bibnamefont
  {Sokolov}}, \bibinfo {author} {\bibfnamefont {R.~E.}\ \bibnamefont
  {Goldstein}}, \bibinfo {author} {\bibfnamefont {F.~I.}\ \bibnamefont
  {Feldchtein}}, \ and\ \bibinfo {author} {\bibfnamefont {I.~S.}\ \bibnamefont
  {Aranson}},\ }\href@noop {} {\bibfield  {journal} {\bibinfo  {journal} {Phys.
  Rev. E.}\ }\textbf {\bibinfo {volume} {80}},\ \bibinfo {pages} {031903}
  (\bibinfo {year} {2009})}\BibitemShut {NoStop}%
\bibitem [{\citenamefont {Ben-Jacob}\ and\ \citenamefont
  {Cohen}(2000)}]{BenJacob2000}%
  \BibitemOpen
  \bibfield  {author} {\bibinfo {author} {\bibfnamefont {E.}~\bibnamefont
  {Ben-Jacob}}\ and\ \bibinfo {author} {\bibfnamefont {I.}~\bibnamefont
  {Cohen}},\ }\href@noop {} {\bibfield  {journal} {\bibinfo  {journal} {Adv. in
  Phys.}\ }\textbf {\bibinfo {volume} {49}},\ \bibinfo {pages} {395} (\bibinfo
  {year} {2000})}\BibitemShut {NoStop}%
\bibitem [{\citenamefont {Yang}\ \emph {et~al.}(2008)\citenamefont {Yang},
  \citenamefont {Elgeti},\ and\ \citenamefont {Gompper}}]{Yang2008}%
  \BibitemOpen
  \bibfield  {author} {\bibinfo {author} {\bibfnamefont {Y.~Z.}\ \bibnamefont
  {Yang}}, \bibinfo {author} {\bibfnamefont {J.}~\bibnamefont {Elgeti}}, \ and\
  \bibinfo {author} {\bibfnamefont {G.}~\bibnamefont {Gompper}},\ }\href@noop
  {} {\bibfield  {journal} {\bibinfo  {journal} {Phys. Rev. E}\ }\textbf
  {\bibinfo {volume} {78}},\ \bibinfo {pages} {061903} (\bibinfo {year}
  {2008})}\BibitemShut {NoStop}%
\bibitem [{\citenamefont {Riedel}\ \emph {et~al.}(2005)\citenamefont {Riedel},
  \citenamefont {Kruse},\ and\ \citenamefont {Howard}}]{Riedel2005}%
  \BibitemOpen
  \bibfield  {author} {\bibinfo {author} {\bibfnamefont {I.~H.}\ \bibnamefont
  {Riedel}}, \bibinfo {author} {\bibfnamefont {K.}~\bibnamefont {Kruse}}, \
  and\ \bibinfo {author} {\bibfnamefont {J.}~\bibnamefont {Howard}},\
  }\href@noop {} {\bibfield  {journal} {\bibinfo  {journal} {Science}\ }\textbf
  {\bibinfo {volume} {309}},\ \bibinfo {pages} {300} (\bibinfo {year}
  {2005})}\BibitemShut {NoStop}%
\bibitem [{\citenamefont {Redner}\ \emph {et~al.}(2013)\citenamefont {Redner},
  \citenamefont {Hagan},\ and\ \citenamefont {Baskaran}}]{Redner2013}%
  \BibitemOpen
  \bibfield  {author} {\bibinfo {author} {\bibfnamefont {G.~S.}\ \bibnamefont
  {Redner}}, \bibinfo {author} {\bibfnamefont {M.~F.}\ \bibnamefont {Hagan}}, \
  and\ \bibinfo {author} {\bibfnamefont {A.}~\bibnamefont {Baskaran}},\
  }\href@noop {} {\bibfield  {journal} {\bibinfo  {journal} {Phys. Rev. Lett.}\
  }\textbf {\bibinfo {volume} {110}},\ \bibinfo {pages} {055701} (\bibinfo
  {year} {2013})}\BibitemShut {NoStop}%
\bibitem [{\citenamefont {Aranson}\ and\ \citenamefont
  {Tsimring}(2006)}]{Aranson2006}%
  \BibitemOpen
  \bibfield  {author} {\bibinfo {author} {\bibfnamefont {I.~S.}\ \bibnamefont
  {Aranson}}\ and\ \bibinfo {author} {\bibfnamefont {L.~S.}\ \bibnamefont
  {Tsimring}},\ }\href@noop {} {\bibfield  {journal} {\bibinfo  {journal} {Rev.
  of Mod. Phys.}\ }\textbf {\bibinfo {volume} {78}},\ \bibinfo {pages} {641}
  (\bibinfo {year} {2006})}\BibitemShut {NoStop}%
\bibitem [{\citenamefont {Dunkel}\ \emph {et~al.}(2013)\citenamefont {Dunkel},
  \citenamefont {Heidenreich}, \citenamefont {Drescher}, \citenamefont
  {Wensink}, \citenamefont {B{\"ar}},\ and\ \citenamefont
  {Goldstein}}]{Goldstein2013}%
  \BibitemOpen
  \bibfield  {author} {\bibinfo {author} {\bibfnamefont {J.}~\bibnamefont
  {Dunkel}}, \bibinfo {author} {\bibfnamefont {S.}~\bibnamefont {Heidenreich}},
  \bibinfo {author} {\bibfnamefont {K.}~\bibnamefont {Drescher}}, \bibinfo
  {author} {\bibfnamefont {H.}~\bibnamefont {Wensink}}, \bibinfo {author}
  {\bibfnamefont {M.}~\bibnamefont {B{\"ar}}}, \ and\ \bibinfo {author}
  {\bibfnamefont {R.~E.}\ \bibnamefont {Goldstein}},\ }\href@noop {} {\bibfield
   {journal} {\bibinfo  {journal} {Phys. Rev. Lett.}\ }\textbf {\bibinfo
  {volume} {110}},\ \bibinfo {pages} {228102} (\bibinfo {year}
  {2013})}\BibitemShut {NoStop}%
\bibitem [{\citenamefont {J{\"a}ger}\ \emph {et~al.}(2013)\citenamefont
  {J{\"a}ger}, \citenamefont {Stark},\ and\ \citenamefont
  {Klapp}}]{Jager2012_2}%
  \BibitemOpen
  \bibfield  {author} {\bibinfo {author} {\bibfnamefont {S.}~\bibnamefont
  {J{\"a}ger}}, \bibinfo {author} {\bibfnamefont {H.}~\bibnamefont {Stark}}, \
  and\ \bibinfo {author} {\bibfnamefont {S.~H.~L.}\ \bibnamefont {Klapp}},\
  }\href@noop {} {\bibfield  {journal} {\bibinfo  {journal} {J. Phys.: Condens.
  Matter}\ }\textbf {\bibinfo {volume} {25}},\ \bibinfo {pages} {195104}
  (\bibinfo {year} {2013})}\BibitemShut {NoStop}%
\bibitem [{\citenamefont {Qian}\ \emph {et~al.}(2009)\citenamefont {Qian},
  \citenamefont {Jiang}, \citenamefont {Gagnon}, \citenamefont {Breuer},\ and\
  \citenamefont {Powers}}]{Qian2009}%
  \BibitemOpen
  \bibfield  {author} {\bibinfo {author} {\bibfnamefont {B.}~\bibnamefont
  {Qian}}, \bibinfo {author} {\bibfnamefont {H.}~\bibnamefont {Jiang}},
  \bibinfo {author} {\bibfnamefont {D.~A.}\ \bibnamefont {Gagnon}}, \bibinfo
  {author} {\bibfnamefont {K.~S.}\ \bibnamefont {Breuer}}, \ and\ \bibinfo
  {author} {\bibfnamefont {T.~R.}\ \bibnamefont {Powers}},\ }\href@noop {}
  {\bibfield  {journal} {\bibinfo  {journal} {Phys. Rev. E}\ }\textbf {\bibinfo
  {volume} {80}},\ \bibinfo {pages} {061919} (\bibinfo {year}
  {2009})}\BibitemShut {NoStop}%
\bibitem [{\citenamefont {Palacci}\ \emph {et~al.}(2013)\citenamefont
  {Palacci}, \citenamefont {Sacanna}, \citenamefont {Steinberg}, \citenamefont
  {Pine},\ and\ \citenamefont {Chaikin}}]{Palacci2013}%
  \BibitemOpen
  \bibfield  {author} {\bibinfo {author} {\bibfnamefont {J.}~\bibnamefont
  {Palacci}}, \bibinfo {author} {\bibfnamefont {S.}~\bibnamefont {Sacanna}},
  \bibinfo {author} {\bibfnamefont {A.~P.}\ \bibnamefont {Steinberg}}, \bibinfo
  {author} {\bibfnamefont {D.~J.}\ \bibnamefont {Pine}}, \ and\ \bibinfo
  {author} {\bibfnamefont {P.~M.}\ \bibnamefont {Chaikin}},\ }\href@noop {}
  {\bibfield  {journal} {\bibinfo  {journal} {Science}\ }\textbf {\bibinfo
  {volume} {339}},\ \bibinfo {pages} {936} (\bibinfo {year}
  {2013})}\BibitemShut {NoStop}%
\bibitem [{\citenamefont {Takagi}\ \emph {et~al.}(2013)\citenamefont {Takagi},
  \citenamefont {Braunschweig}, \citenamefont {Zhang},\ and\ \citenamefont
  {Shelley}}]{Takagi2013}%
  \BibitemOpen
  \bibfield  {author} {\bibinfo {author} {\bibfnamefont {D.}~\bibnamefont
  {Takagi}}, \bibinfo {author} {\bibfnamefont {A.~B.}\ \bibnamefont
  {Braunschweig}}, \bibinfo {author} {\bibfnamefont {J.}~\bibnamefont {Zhang}},
  \ and\ \bibinfo {author} {\bibfnamefont {M.~J.}\ \bibnamefont {Shelley}},\
  }\href@noop {} {\bibfield  {journal} {\bibinfo  {journal} {Phys. Rev. Lett.}\
  }\textbf {\bibinfo {volume} {110}},\ \bibinfo {pages} {038301} (\bibinfo
  {year} {2013})}\BibitemShut {NoStop}%
\bibitem [{\citenamefont {Tierno}\ \emph {et~al.}(2007)\citenamefont {Tierno},
  \citenamefont {Muruganathan},\ and\ \citenamefont {Fischer}}]{Tierno2007}%
  \BibitemOpen
  \bibfield  {author} {\bibinfo {author} {\bibfnamefont {P.}~\bibnamefont
  {Tierno}}, \bibinfo {author} {\bibfnamefont {R.}~\bibnamefont
  {Muruganathan}}, \ and\ \bibinfo {author} {\bibfnamefont {T.~M.}\
  \bibnamefont {Fischer}},\ }\href@noop {} {\bibfield  {journal} {\bibinfo
  {journal} {Phys. Rev. Lett.}\ }\textbf {\bibinfo {volume} {98}},\ \bibinfo
  {pages} {028301} (\bibinfo {year} {2007})}\BibitemShut {NoStop}%
\bibitem [{\citenamefont {Petousis}\ \emph {et~al.}(2007)\citenamefont
  {Petousis}, \citenamefont {Homburg}, \citenamefont {Derks},\ and\
  \citenamefont {Dietzel}}]{Petousis2007}%
  \BibitemOpen
  \bibfield  {author} {\bibinfo {author} {\bibfnamefont {I.}~\bibnamefont
  {Petousis}}, \bibinfo {author} {\bibfnamefont {E.}~\bibnamefont {Homburg}},
  \bibinfo {author} {\bibfnamefont {R.}~\bibnamefont {Derks}}, \ and\ \bibinfo
  {author} {\bibfnamefont {A.}~\bibnamefont {Dietzel}},\ }\href@noop {}
  {\bibfield  {journal} {\bibinfo  {journal} {Lab Chip}\ }\textbf {\bibinfo
  {volume} {7}},\ \bibinfo {pages} {1746} (\bibinfo {year} {2007})}\BibitemShut
  {NoStop}%
\bibitem [{\citenamefont {Biswal}\ and\ \citenamefont
  {Gast}(2004)}]{Biswal2004}%
  \BibitemOpen
  \bibfield  {author} {\bibinfo {author} {\bibfnamefont {S.~L.}\ \bibnamefont
  {Biswal}}\ and\ \bibinfo {author} {\bibfnamefont {A.~P.}\ \bibnamefont
  {Gast}},\ }\href@noop {} {\bibfield  {journal} {\bibinfo  {journal} {Phys.
  Rev. E}\ }\textbf {\bibinfo {volume} {69}},\ \bibinfo {pages} {041406}
  (\bibinfo {year} {2004})}\BibitemShut {NoStop}%
\bibitem [{\citenamefont {Vuppu}\ \emph {et~al.}(2003)\citenamefont {Vuppu},
  \citenamefont {Garcia},\ and\ \citenamefont {Hayes}}]{Vuppu2003}%
  \BibitemOpen
  \bibfield  {author} {\bibinfo {author} {\bibfnamefont {A.~K.}\ \bibnamefont
  {Vuppu}}, \bibinfo {author} {\bibfnamefont {A.~A.}\ \bibnamefont {Garcia}}, \
  and\ \bibinfo {author} {\bibfnamefont {M.~A.}\ \bibnamefont {Hayes}},\
  }\href@noop {} {\bibfield  {journal} {\bibinfo  {journal} {Langmuir}\
  }\textbf {\bibinfo {volume} {19}},\ \bibinfo {pages} {8646} (\bibinfo {year}
  {2003})}\BibitemShut {NoStop}%
\bibitem [{\citenamefont {Furst}\ and\ \citenamefont {Gast}(1999)}]{Furst1999}%
  \BibitemOpen
  \bibfield  {author} {\bibinfo {author} {\bibfnamefont {E.~M.}\ \bibnamefont
  {Furst}}\ and\ \bibinfo {author} {\bibfnamefont {A.~P.}\ \bibnamefont
  {Gast}},\ }\href@noop {} {\bibfield  {journal} {\bibinfo  {journal} {Phys.
  Rev. Lett.}\ }\textbf {\bibinfo {volume} {82}},\ \bibinfo {pages} {4130}
  (\bibinfo {year} {1999})}\BibitemShut {NoStop}%
\bibitem [{\citenamefont {Helgesen}\ \emph {et~al.}(1988)\citenamefont
  {Helgesen}, \citenamefont {Skjeltorp}, \citenamefont {Mors}, \citenamefont
  {Botet},\ and\ \citenamefont {Jullien}}]{Helgesen1988}%
  \BibitemOpen
  \bibfield  {author} {\bibinfo {author} {\bibfnamefont {G.}~\bibnamefont
  {Helgesen}}, \bibinfo {author} {\bibfnamefont {A.~T.}\ \bibnamefont
  {Skjeltorp}}, \bibinfo {author} {\bibfnamefont {P.~M.}\ \bibnamefont {Mors}},
  \bibinfo {author} {\bibfnamefont {R.}~\bibnamefont {Botet}}, \ and\ \bibinfo
  {author} {\bibfnamefont {R.}~\bibnamefont {Jullien}},\ }\href@noop {}
  {\bibfield  {journal} {\bibinfo  {journal} {Phys. Rev. Lett.}\ }\textbf
  {\bibinfo {volume} {61}},\ \bibinfo {pages} {1736} (\bibinfo {year}
  {1988})}\BibitemShut {NoStop}%
\bibitem [{\citenamefont {J{\"a}ger}\ \emph {et~al.}(2012)\citenamefont
  {J{\"a}ger}, \citenamefont {Schmidle},\ and\ \citenamefont
  {Klapp}}]{Jager2012}%
  \BibitemOpen
  \bibfield  {author} {\bibinfo {author} {\bibfnamefont {S.}~\bibnamefont
  {J{\"a}ger}}, \bibinfo {author} {\bibfnamefont {H.}~\bibnamefont {Schmidle}},
  \ and\ \bibinfo {author} {\bibfnamefont {S.~H.~L.}\ \bibnamefont {Klapp}},\
  }\href@noop {} {\bibfield  {journal} {\bibinfo  {journal} {Phys. Rev. E}\
  }\textbf {\bibinfo {volume} {86}},\ \bibinfo {pages} {011402} (\bibinfo
  {year} {2012})}\BibitemShut {NoStop}%
\bibitem [{\citenamefont {Regtmeier}\ \emph {et~al.}(2012)\citenamefont
  {Regtmeier}, \citenamefont {Wittbracht}, \citenamefont {Rempel},
  \citenamefont {Mill}, \citenamefont {Peter}, \citenamefont {Weddemann},
  \citenamefont {Mattay},\ and\ \citenamefont {H{\"u}tten}}]{Regtmeier2012}%
  \BibitemOpen
  \bibfield  {author} {\bibinfo {author} {\bibfnamefont {A.}~\bibnamefont
  {Regtmeier}}, \bibinfo {author} {\bibfnamefont {F.}~\bibnamefont
  {Wittbracht}}, \bibinfo {author} {\bibfnamefont {T.}~\bibnamefont {Rempel}},
  \bibinfo {author} {\bibfnamefont {N.}~\bibnamefont {Mill}}, \bibinfo {author}
  {\bibfnamefont {M.}~\bibnamefont {Peter}}, \bibinfo {author} {\bibfnamefont
  {A.}~\bibnamefont {Weddemann}}, \bibinfo {author} {\bibfnamefont
  {J.}~\bibnamefont {Mattay}}, \ and\ \bibinfo {author} {\bibfnamefont
  {A.}~\bibnamefont {H{\"u}tten}},\ }\href@noop {} {\bibfield  {journal}
  {\bibinfo  {journal} {J. Nanopart. Res.}\ }\textbf {\bibinfo {volume} {14}},\
  \bibinfo {pages} {098103} (\bibinfo {year} {2012})}\BibitemShut {NoStop}%
\bibitem [{\citenamefont {Wittbracht}\ \emph {et~al.}(2011)\citenamefont
  {Wittbracht}, \citenamefont {Eickenberg}, \citenamefont {Weddemann},\ and\
  \citenamefont {H{\"u}tten}}]{Wittbracht2011}%
  \BibitemOpen
  \bibfield  {author} {\bibinfo {author} {\bibfnamefont {F.}~\bibnamefont
  {Wittbracht}}, \bibinfo {author} {\bibfnamefont {B.}~\bibnamefont
  {Eickenberg}}, \bibinfo {author} {\bibfnamefont {A.}~\bibnamefont
  {Weddemann}}, \ and\ \bibinfo {author} {\bibfnamefont {A.}~\bibnamefont
  {H{\"u}tten}},\ }\href@noop {} {\bibfield  {journal} {\bibinfo  {journal} {J.
  Appl. Phys.}\ }\textbf {\bibinfo {volume} {109}},\ \bibinfo {pages} {114503}
  (\bibinfo {year} {2011})}\BibitemShut {NoStop}%
\bibitem [{\citenamefont {Derks}(2010)}]{Derks2010}%
  \BibitemOpen
  \bibfield  {author} {\bibinfo {author} {\bibfnamefont {R.}~\bibnamefont
  {Derks}},\ }\emph {\bibinfo {title} {Particle Dynamics in Magneto-Fluidic
  Microsystems}},\ \href@noop {} {\bibinfo {type} {{Ph.D.} thesis}},\ \bibinfo
  {school} {Eindhoven University} (\bibinfo {year} {2010})\BibitemShut
  {NoStop}%
\bibitem [{\citenamefont {Snoswell}\ \emph {et~al.}(2006)\citenamefont
  {Snoswell}, \citenamefont {Bower}, \citenamefont {Ivanov}, \citenamefont
  {Cryan}, \citenamefont {Rarity},\ and\ \citenamefont
  {Vincent}}]{Snoswell2006}%
  \BibitemOpen
  \bibfield  {author} {\bibinfo {author} {\bibfnamefont {D.~R.~E.}\
  \bibnamefont {Snoswell}}, \bibinfo {author} {\bibfnamefont {C.~L.}\
  \bibnamefont {Bower}}, \bibinfo {author} {\bibfnamefont {P.}~\bibnamefont
  {Ivanov}}, \bibinfo {author} {\bibfnamefont {M.~J.}\ \bibnamefont {Cryan}},
  \bibinfo {author} {\bibfnamefont {J.~G.}\ \bibnamefont {Rarity}}, \ and\
  \bibinfo {author} {\bibfnamefont {B.}~\bibnamefont {Vincent}},\ }\href@noop
  {} {\bibfield  {journal} {\bibinfo  {journal} {New J. Phys.}\ }\textbf
  {\bibinfo {volume} {8}},\ \bibinfo {pages} {267} (\bibinfo {year}
  {2006})}\BibitemShut {NoStop}%
\bibitem [{Sup()}]{SuppMat}%
  \BibitemOpen
  \href@noop {} {\bibinfo  {journal} {See Supplemental Material at [URL] for
  videos of self-assembling monolayers of paramagnetic particles.}\
  }\BibitemShut {NoStop}%
\bibitem [{\citenamefont {Dinsmore}\ \emph {et~al.}(2001)\citenamefont
  {Dinsmore}, \citenamefont {Weeks}, \citenamefont {Prasad}, \citenamefont
  {Levitt},\ and\ \citenamefont {Weitz}}]{Dinsmore2001}%
  \BibitemOpen
\bibfield  {journal} {  }\bibfield  {author} {\bibinfo {author} {\bibfnamefont
  {A.~D.}\ \bibnamefont {Dinsmore}}, \bibinfo {author} {\bibfnamefont {E.~R.}\
  \bibnamefont {Weeks}}, \bibinfo {author} {\bibfnamefont {V.}~\bibnamefont
  {Prasad}}, \bibinfo {author} {\bibfnamefont {A.~C.}\ \bibnamefont {Levitt}},
  \ and\ \bibinfo {author} {\bibfnamefont {D.~A.}\ \bibnamefont {Weitz}},\
  }\href@noop {} {\bibfield  {journal} {\bibinfo  {journal} {Appl. Optics}\
  }\textbf {\bibinfo {volume} {40}},\ \bibinfo {pages} {4152} (\bibinfo {year}
  {2001})}\BibitemShut {NoStop}%
\bibitem [{\citenamefont {Gonz$\acute{\mbox{a}}$lez}\ \emph
  {et~al.}(1999)\citenamefont {Gonz$\acute{\mbox{a}}$lez}, \citenamefont
  {Lach-Hab},\ and\ \citenamefont {Blaisten-Barojas}}]{Gonzalez1999}%
  \BibitemOpen
  \bibfield  {author} {\bibinfo {author} {\bibfnamefont {A.~E.}\ \bibnamefont
  {Gonz$\acute{\mbox{a}}$lez}}, \bibinfo {author} {\bibfnamefont
  {M.}~\bibnamefont {Lach-Hab}}, \ and\ \bibinfo {author} {\bibfnamefont
  {E.}~\bibnamefont {Blaisten-Barojas}},\ }\href@noop {} {\bibfield  {journal}
  {\bibinfo  {journal} {J. of Sol-Gel Sci. and Tech.}\ }\textbf {\bibinfo
  {volume} {15}},\ \bibinfo {pages} {119} (\bibinfo {year} {1999})}\BibitemShut
  {NoStop}%
\bibitem [{\citenamefont {Wittbracht}\ \emph {et~al.}(2012)\citenamefont
  {Wittbracht}, \citenamefont {Weddemann}, \citenamefont {Eickenberg},\ and\
  \citenamefont {H{\"u}tten}}]{Wittbracht2012}%
  \BibitemOpen
  \bibfield  {author} {\bibinfo {author} {\bibfnamefont {F.}~\bibnamefont
  {Wittbracht}}, \bibinfo {author} {\bibfnamefont {A.}~\bibnamefont
  {Weddemann}}, \bibinfo {author} {\bibfnamefont {B.}~\bibnamefont
  {Eickenberg}}, \ and\ \bibinfo {author} {\bibfnamefont {A.}~\bibnamefont
  {H{\"u}tten}},\ }\href@noop {} {\bibfield  {journal} {\bibinfo  {journal}
  {Microfluid Nanofluid}\ }\textbf {\bibinfo {volume} {13}},\ \bibinfo {pages}
  {543} (\bibinfo {year} {2012})}\BibitemShut {NoStop}%
\bibitem [{\citenamefont {Calder{\'o}n}\ and\ \citenamefont
  {Melle}(2002)}]{Calderon2002}%
  \BibitemOpen
  \bibfield  {author} {\bibinfo {author} {\bibfnamefont {O.~G.}\ \bibnamefont
  {Calder{\'o}n}}\ and\ \bibinfo {author} {\bibfnamefont {S.}~\bibnamefont
  {Melle}},\ }\href@noop {} {\bibfield  {journal} {\bibinfo  {journal} {J.
  Phys. D: Appl. Phys.}\ }\textbf {\bibinfo {volume} {35}},\ \bibinfo {pages}
  {2492} (\bibinfo {year} {2002})}\BibitemShut {NoStop}%
\bibitem [{\citenamefont {Tailleur}\ and\ \citenamefont
  {Cates}(2008)}]{Tailleur2008}%
  \BibitemOpen
  \bibfield  {author} {\bibinfo {author} {\bibfnamefont {J.}~\bibnamefont
  {Tailleur}}\ and\ \bibinfo {author} {\bibfnamefont {M.~E.}\ \bibnamefont
  {Cates}},\ }\href@noop {} {\bibfield  {journal} {\bibinfo  {journal} {Phys.
  Rev. Lett.}\ }\textbf {\bibinfo {volume} {100}},\ \bibinfo {pages} {218103}
  (\bibinfo {year} {2008})}\BibitemShut {NoStop}%
\bibitem [{\citenamefont {Reichert}\ and\ \citenamefont
  {Stark}(2005)}]{Reichert2005}%
  \BibitemOpen
  \bibfield  {author} {\bibinfo {author} {\bibfnamefont {M.}~\bibnamefont
  {Reichert}}\ and\ \bibinfo {author} {\bibfnamefont {H.}~\bibnamefont
  {Stark}},\ }\href@noop {} {\bibfield  {journal} {\bibinfo  {journal} {Eur.
  Phys. J. E}\ }\textbf {\bibinfo {volume} {17}},\ \bibinfo {pages} {493}
  (\bibinfo {year} {2005})}\BibitemShut {NoStop}%
\end{thebibliography}%

\end{document}